# Nature of non-magnetic strongly-correlated state in δ-plutonium

L.V. Pourovskii[1], M.I. Katsnelson[1], A.I. Lichtenstein[2], L. Havela[3], T. Gouder[4], F. Wastin[4,] A.B. Shick[5], V. Drchal[5] and G.H. Lander[4]

[1]*Institute for Molecules and Materials, Radboud University of Nijmegen, NL-6525 ED Nijmegen, The Netherlands*

[2]*Institut für Theoretische Physik, Universität Hamburg, 20355 Hamburg, Germany*

[3]*Charles University, Faculty of Mathematics and Physics, Department of Electronic Structures, 121 16 Prague 2, Czech Republic*

[4]*European Commission, JRC, Institute for Transuranium Elements, Karlsruhe, P.O. Box 2340, 76 125 Karlsruhe, Germany*

[5]*Institute of Physics, ASCR, 182 21 Prague 8, Czech Republic*

**The solid-state properties of most of elements are now well understood on the basis of quantum physics – with few exceptions, notably the element number 94, plutonium. For Pu, difficulties have been known for many years, hence the large number of studies, especially theoretical, of this mysterious element. Plutonium has six crystalline phases at ambient pressure, some of which are separated by unusual phase transitions (with large discontinuities in volume), exhibit negative thermal expansion coefficients, or form exotic low-symmetry structures[1,2]. The main challenge to explain these anomalous properties is that the characteristic ingredient of actinides, the 5*f* electronic states, are in the cross-over regime between the localized and delocalized (itinerant) behaviour in Pu[3,4]. The early part of the actinide series with the 5*f* states being itinerant, i.e. part of the metallic bond, culminates with Pu; starting with Am (Z = 95), the 5*f* states are localized, non-bonding, and resemble the 4*f* states in lanthanides. Both itinerant and localized regimes are well covered by existing theories, but they cannot be simply interpolated due to the importance of many-body electron correlations[5,6]. The fundamental problem for Pu is that theories of strongly correlated systems exhibit local magnetic moments (ordered or disordered), whereas experimental data in Pu demonstrate unambiguously their absence[7]. Standard band-structure calculations predict strong magnetism for different phases of Pu[8,4], and local moments appear as a crucial ingredient for an adequate description of equilibrium lattice constants and bulk moduli[9,10]. Here we demonstrate the solution to this problem by including dynamical correlation effects; the spin-orbital fluctuations involving $f^5$ and $f^6$ configurations suppress the magnetism of plutonium, and result in a many-electron resonance near the Fermi level, in an agreement with high-resolution photoelectron spectra.**

The basic features of light actinides are fully explained by relativistic band theories[8], which correctly describe features such as the low symmetry crystal structures and lack of magnetic order. To some extent this is true also for the Pu ground-state (α-Pu), which crystallizes in a monoclinic structure. Its small atomic volume implies the bonding character of the 5*f* states.

However, as is well known,[1,2] Pu exhibits five more allotropic phases before melting, and all have much larger volumes. The most enigmatic is δ-Pu phase with the simple *fcc* crystal structure, which is also the most technologically important one. The volume of this phase is 25% greater than α-Pu and already half way towards the volume of Am metal, which has localized 5*f* states at ambient conditions but can pass to the mixed valence state at high pressures.[9]

The atomic volume, being sensitive to the spatial extent of the 5*f* states (which screen the attractive potential of the atomic core acting on peripheral valence electrons) can be taken as the prime indicator of the character of electronic states. The enormous increase of the volume (by 25% between α-and δ-Pu) leads to a dramatic narrowing of the 5*f* band [4]. Since the bands are already relatively narrow, within the framework of the density functional theory (DFT) with the local density (LDA) or generalized gradient (GGA) approximations, the further narrowing in energy space leads inexorably to predictions of the presence of magnetic moments, either static or dynamic, in the δ-phase of Pu. However, the ensemble of experiments, performed over many years, shows unambiguously that there is *no* evidence for either static or dynamic magnetic moments in δ-Pu.[7] The resulting controversy between experiment and theory is difficult to reconcile, because, although a partial cancellation of spin and orbital moments can be expected,[10] the Curie-Weiss behaviour should be observed in the paramagnetic region, but the observed magnetic susceptibility remains practically temperature independent (δ-Pu exists between 592 and 724 K but can be stabilized down to $T \to 0$ K by the addition of a small percentage of dopants). Magnetic ordering in δ-Pu is further excluded by NMR [11] and neutron-scattering experiments.[7] In particular, the latter reveal no magnetic Bragg intensity and all the inelastic scattering in the energy-transfer window 0-20 meV can be attributed to processes involving phonons. More recently, detailed muon experiments[12] further exclude any static or dynamic magnetic moment on the scale of $10^{-3}$ $\mu_B$.

We are hence forced to consider alternative methods to conventional DFT for an understanding of the features such as the large atomic volume, non-magnetic ground state, and relatively high quasiparticle density-of-states at the Fermi level $E_F$, deduced from the electronic specific heat coefficient $\gamma = 52$ mJ/mol K$^2$ ([1,7]), that characterises the ground-state of δ-Pu. Important direct information is provided by photoelectron spectroscopy. High-resolution ultraviolet photoelectron spectroscopy (UPS) reveals a high density of quasi-particle states at, or near to, $E_F$, (**Fig. 1**) and a narrow satellite maximum at 0.85 eV binding energy (BE, related to $E_F$).[13-15] A detailed analysis resolves also a weaker feature at 0.5 eV. Using the variation with energy of the photo-excitation cross-section for different electronic states, the 5*f* character is found to dominate all three features, and similar spectral features are found, as shown in Fig. 1, for many systems containing Pu. This triplet in the spectral response is gradually suppressed for ultra-thin Pu slabs, and a broader 5*f* feature starts to emerge and shift to higher BE, being finally located at about 1.6 eV in the limit of one atomic layer[15].

A rigorous description of the ground-state electron configuration is of crucial importance to treat the physical and chemical properties of actinide metals. The "Mixed level" approach[16] with separation of *f* – electrons into itinerant and core states, with subsequent calculations of the total energy in standard LDA or GGA, leads to the conclusion that the minimal energy for δ-Pu corresponds to a $f^4$ configuration. On the other hand, opposite to recent magnetic

GGA attempt,[17] the so-called "*LDA+U*" method[18] demonstrates a possible stable *non-magnetic* configuration, derived from the $f^6$ state hybridized with conduction electrons.[19] The key ingredient is a large spin-orbit splitting with completely filled $f_{5/2}$ band and an empty $f_{7/2}$ band state. However, this picture concludes that there is no, or very little, *f* – electron spectral density near the Fermi energy, which clearly contradicts the experimental evidence from both UPS (see Fig. 1) and also from that inferred from the high electronic coefficient in the specific heat.[7] In reality is not correct to consider all *f*–electron configurations as equally localized. The well-known phenomenon of a wave-function collapse in a two-well potential for *f*-electrons[20] may be a key for understanding the problem. It was shown[21] that in the Pu atom an essential transfer of the electron density from the internal to the external well takes place exactly at the transition from the $f^5$ to $f^6$ configuration; and this can lead to a partial delocalization of *f*–electrons in the latter configuration. In part, this recalls a mechanism proposed by Johansson[22] for the α–γ transition in cerium and for the intermediate valence of α-cerium. As shown below, by taking into account correlation effects it is possible to describe correctly the electronic structure of δ-Pu. A part of the *f*–spectral density moves to the Fermi energy, while the total state remains non-magnetic.

In the *LDA+U* technique one takes into account only the static part of the full self-energy, the part which usually gives the most important contribution of correlations to the thermodynamic properties of a system, but, as a rule, is not sufficient to reproduce correctly spectral properties of strongly correlated systems. We took into account the static part of the self-energy using the same type of *LDA+U* approach as Shick *et al.*[19] and obtained the dynamical part of the self energy via the dynamical mean-field theory (DMFT)[23,5]. We include the on-shell Coulomb interaction only for Pu *f*-electrons; it can be defined by the "direct" and "exchange" parameters *U* and *J*. In agreement with results of Ref. 19, our LDA+*U* calculations with reasonable values of the parameters[10] *U* = 3 - 4 eV and *J* = 0.7 eV predict a non-magnetic ground state of δ-Pu with both lattice parameter and bulk modulus in good agreement with experiment. The theoretical lattice parameters are 4.58 and 4.78 Å, and theoretical bulk moduli are 442 and 370 kBar for *U* equal to 3 and 4 eV, respectively; the experimental values of the lattice parameter and bulk modulus being 4.64 Å and 300-350 kBar, respectively. We then use the static part of the self-energy obtained within *LDA+U* method and employ the DMFT to compute the dynamical part of self-energy $\Sigma(\omega)$ and local GF $G(\omega)$. The complete description of our LDA+U-DMFT technique can be found elsewhere[24].

The comparison of the density of states (DOS) obtained within the DMFT (**Fig. 2a**) and the *LDA+U* (Ref. 18) methods reveals a profound impact of dynamics of spin-orbital correlations on the electronic structure of δ-Pu. The most important modification is the formation of a high peak just below the Fermi level (0.3 eV BE) due to *f*-electrons; this feature is missing in the DOS derived from *LDA+U* theory,[19] but is observed in photoemission spectra. The remainder of the lower Hubbard band is spread between 3 eV and 1 eV BE in the DMFT DOS, and additional two peaks at 0.55 and 0.8 eV. Such two peaks are also observed in experimental photoemission spectra (at 0.5 eV and 0.85 eV BE). Overall we find good qualitative agreement between the experimental photoemission spectra and the DMFT density of states obtained for the non-magnetic ground state of δ-Pu. Also the most prominent peak in the vicinity of the Fermi level is not at 0.1 eV, but at 0.3 eV BE. The DMFT spectral density (**Fig. 2b**) shows non-trivial dependence of the quasiparticle peak and two low-energy satellites on the quasimomentum in Brillouin zone and can be detected in angle-resolved photoemission experiments.

It is a common habit in the DMFT to interpret the peak near the Fermi energy as a "Kondo resonance".[22] Thus, it appears that such a peak in the DOS is intimately connected with the existence of local magnetic moments, which is not the case for δ-Pu. However, the formation of a many-body peak near the Fermi level is also possible in a more general case for *soft* charge, spin, or orbital fluctuations.[25,26] The energy needed to overcome the spin-orbital gap in the case of δ-Pu is still much smaller than the total $f$ – electron bandwidth, which is enough for a significant mass enhancement near the Fermi surface.[15] Thus, one can say that the many-body feature near $E_F$ is caused by the spin-orbital fluctuations, which are taken into account in the *LDA*+DMFT.

The *f*-state occupation $n_f$ in our calculations is around 5.8 electrons, which means that δ-Pu is an intermediate valence system. However, usually this term is restricted for the situation when the mixing of the two valence states is so slow that fast local-probe methods can distinguish the two boundary states with integer *f*-occupancy. Fingerprints of this two states can be seen in core-level photoemission spectra, but available data for δ-Pu[27] do not seem to confirm this picture. Further, no low-energy quasielastic line, which would reflect the dynamics of the valence fluctuations, was indicated in δ-Pu by neutron inelastic scattering.[7] Therefore, we do not have any direct evidence that δ-Pu is in the valence-fluctuation regime known for some Ce-, Yb-, Sm-, or Eu-based systems. We cannot exclude, however, that the three features close to $E_F$ are related to the $5f^5$ multiplet of final states, originating from the $5f^6$ to $5f^5$ photo-excitation. Such final-state effects were alternatively (besides many-body effects) suggested[28,29] to explain why the valence-band photoemission spectra of PuN and $PuSi_2$ (see Fig. 1) share common features with some other Pu-based systems.

An important experimental test for the theory can be the influence of lattice expansion on the properties of δ-Pu. Systems in which a narrow peak near $E_F$ is a dominant feature will response dramatically even on a small volume expansion as this leads to a further reduction of the bandwidth and an *increase* in $N(E_F)$, reaching eventually the condition for magnetic ordering. If the *f*-states are separated from $E_F$, there is no special reason why the electronic structure should strongly react on the lattice expansion. It is therefore interesting to examine the properties of the materials doped with Am, which maintain the *fcc* structure to high Am concentrations, and considerably expands the lattice (see inset of Fig. 3). The observed magnetic susceptibility, shown in **Fig. 3**, remains on a level lower than that in materials with any type of magnetic order (although, as shown, the Van Vleck susceptibility of pure Am is higher than that for δ-Pu). In addition, the invariant level of Pu-5*f* localization is indicated by the photoelectron spectroscopy.[30] These results exclude that a narrow *5f* band remains located at the Fermi level, and the picture with one-electron states below the Fermi level is strongly preferred. Many-body effects are then responsible for the high density of quasi-particle states near $E_F$ seen in the UPS, as well as in the high $\gamma$-coefficient of specific heat in all these materials.

To conclude, we emphasize that the present calculations, using *ab-initio* dynamical mean-field theory, have been able to resolve the long-standing controversy between theory and experiment in the "simple" face-centered cubic phase of plutonium called δ-Pu. In agreement with experiment, neither static nor dynamical magnetic moments are predicted, but the large peak of *f*-character in the density of states near the Fermi level (sensed in spectroscopic measurements and indirectly by the specific heat) is attributed to spin-orbital fluctuations. The success of this theory on a "simple" element further opens the way for its application to a host of fascinating compounds that challenge condensed-matter theorists.

**Acknowledgements.** This work was partially supported by the EU Research Training Network "Ab-initio Computation of Electronic Properties of *f*-electron Materials" (contract HPRN-CT-2002-00295). The work of L.H. was supported by the Grant Agency of the Czech Republic under the grant No. 202/04/1103.

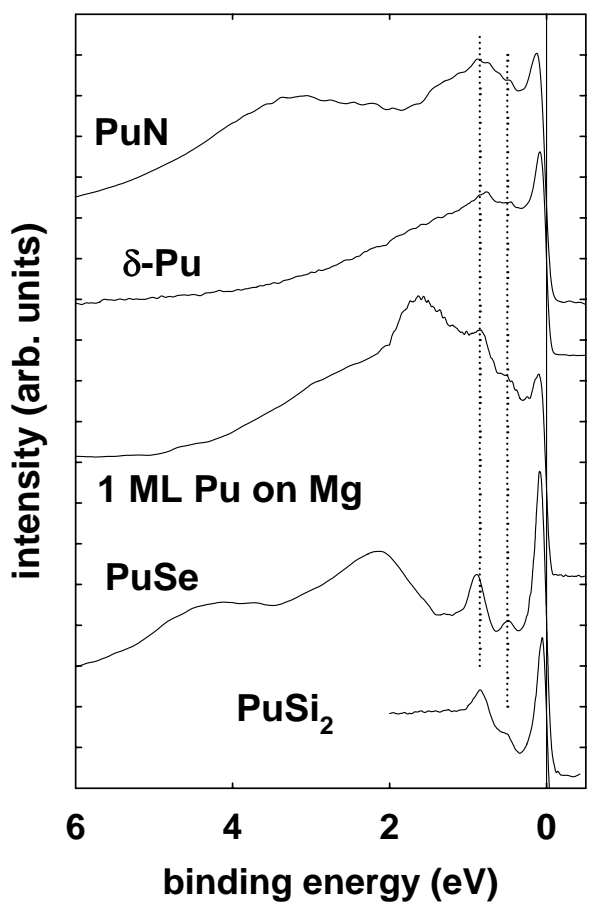

Fig.1: HeII spectra (excitation energy $h\nu$ = 40.81 eV) of selected Pu systems, including 1 monolayer of Pu on Mg. Although the character of individual spectra differs markedly, they all exhibit the triplet of narrow spectral lines at the Fermi level, at 0.5 eV and 0.85 eV. This implies that those lines cannot be related to individual density of one-electron states. (Data compiled from Refs. 28 and 29).

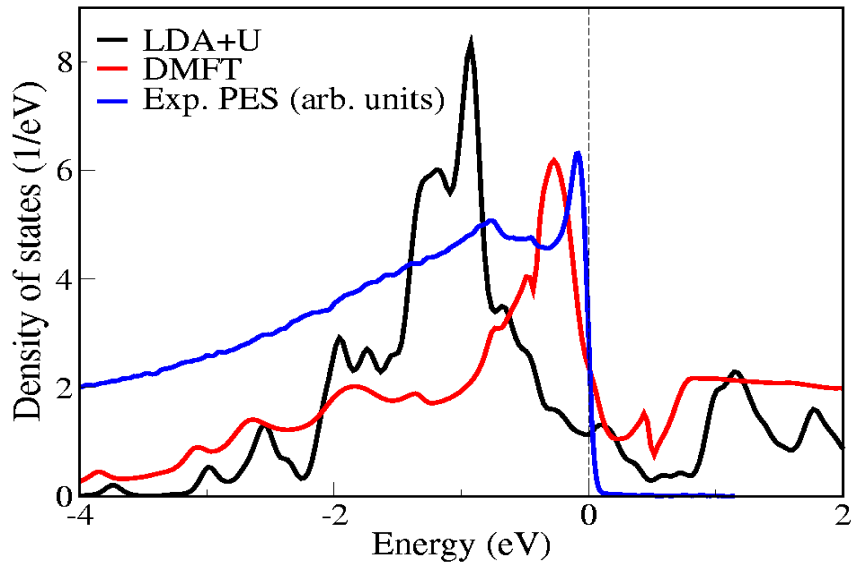

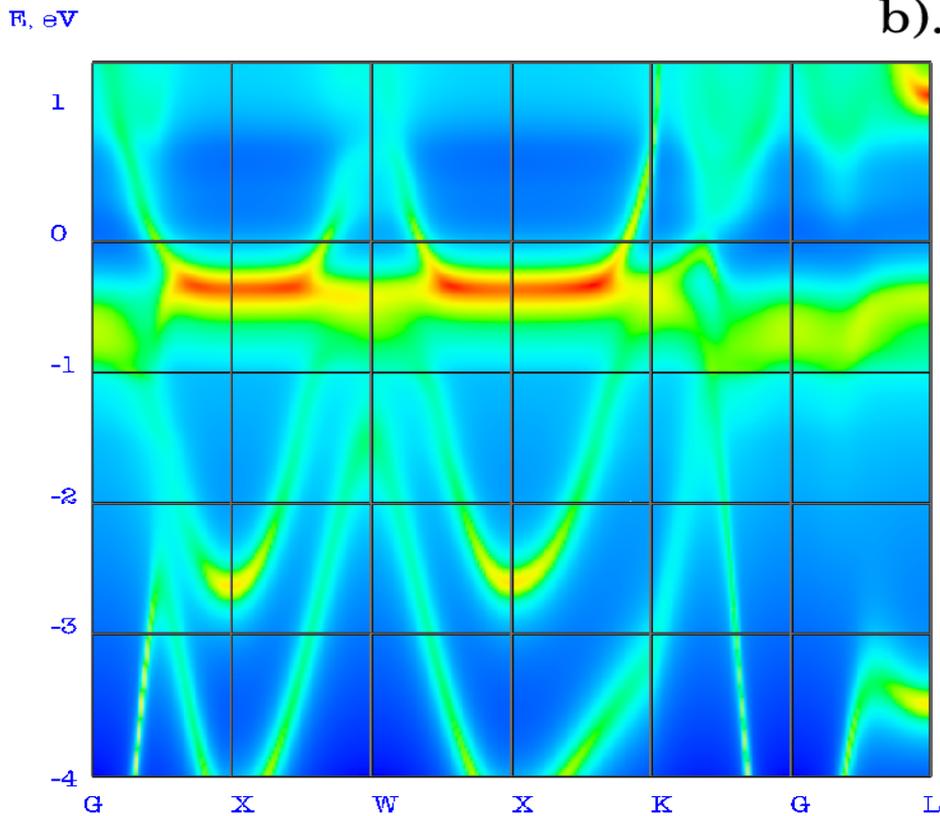

Fig. 2.: a). Density of states as function of energy relative to $E_F$ obtained by the LDA+U (black line) and DMFT (red line) calculations for bulk δ-Pu at the experimental lattice parameter and $U = 3$ eV. The experimental photoemission spectra (in arb. units) is shown by blue line. b). Spectral density for bulk δ-Pu calculated within the DMFT. The sequence of coulors in order of increasing spectral density is blue, green, yellow, and red.

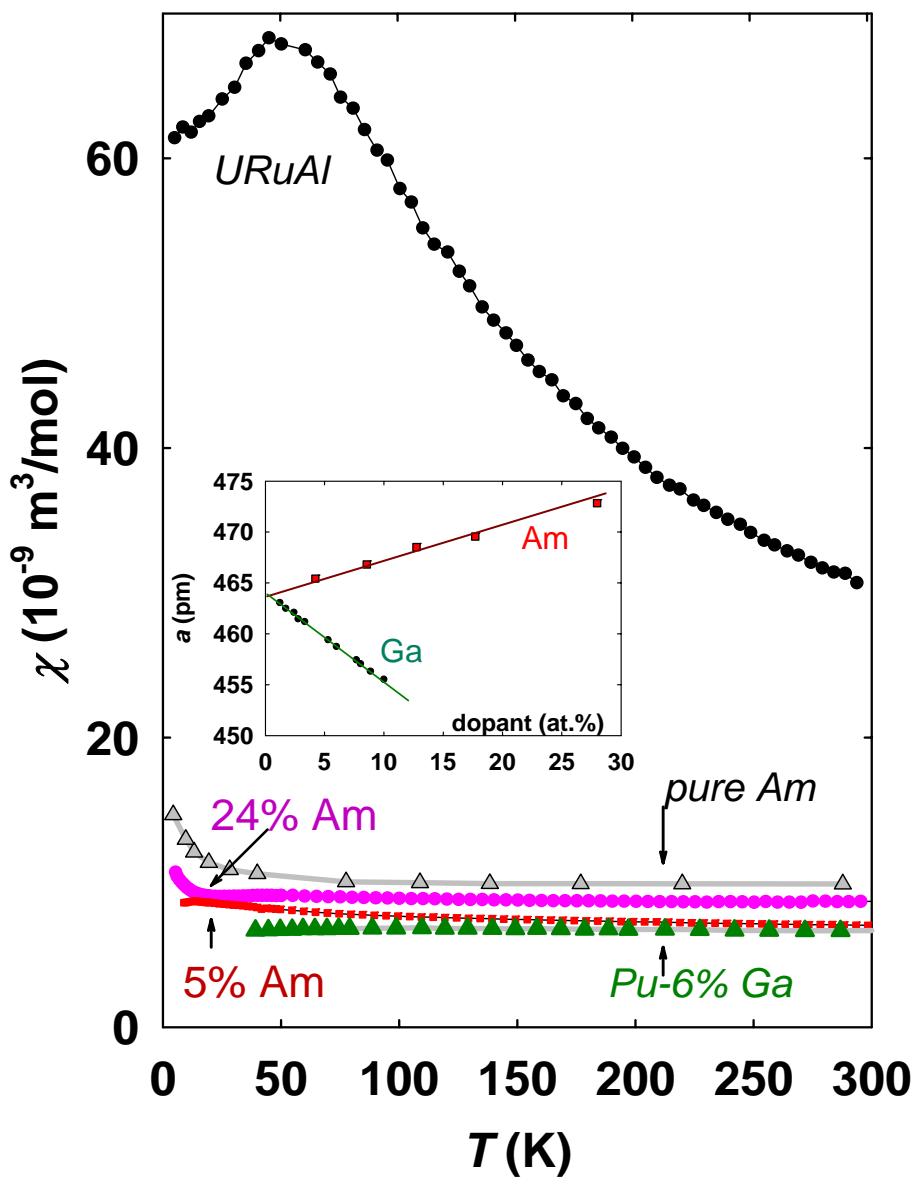

Fig.3: Temperature dependence of magnetic susceptibility for Pu-Am alloys with 5% and 24% Am.[38] The weak upturn at low temperatures may be attributed to Curie terms, arising presumably from a small amount (≈ 0.1 %) of $^{237}$Np, which is a product of α-decay of $^{241}$Am with the half-life of 432 years. The upturn is therefore more pronounced for pure Am. The data are compared with Pu doped by 6% Ga, which also adopts the *fcc* structure, but leads to a volume contraction (concentration variations of the lattice parameter a are displayed in the inset). The example of URuAl, which also does not order magnetically, but exhibits a stronger influence of spin fluctuations, is included to demonstrate the true weakly magnetic nature of all the *fcc* Pu systems discussed in this work.